\documentclass[aps]{revtex4}

\begin{document}

\title{Dynamical symmetry and analytical solutions of the non-autonomous
quantum master equation of the dissipative two-level system:
decoherence of quantum register}
\author{Shun-Jin Wang$^{1,2}$,Jun-Hong An$^2$,Hong-Gang Luo$^2$ and
Cheng-Long Jia$^2 $}
\affiliation{1.Department of Physics of Sichuan University, Chengdu 610065,P.R.China \\
2.Department of Modern Physics of Lanzhou University, Lanzhou
730000,P.R.China}

\begin{abstract}
Based on the non-autonomous quantum master equation, we investigate
the dissipative and decoherence properties of the two-level atom
system interacting with the environment of thermal quantum radiation
fields. For this system, by a novel algebraic dynamic method, the
dynamical symmetry of the system is found, the quantum master
equation is converted into a Schr\"{o}dinger-like equation and the
non-Hermitian rate (quantum Liouville) operator of the master
equation is expressed as a linear function of the dynamical u(2)
generators. Furthermore, the integrability of the non-autonomous
master equation has been proved for the first time. Based on the
time-dependent analytical solutions, the asymptotic behavior of the
solution has been examined and the approach to the equilibrium state
has been proved. Finally, we have studied the decoherence property
of the multiple two-level atom system coupled to the thermal
radiation fields, which are related to the quantum register.
\end{abstract}

\maketitle

\section{Introduction}

The behavior of the quantum dissipative systems interacting with the
background thermal radiation field, is one of the central subjects
in quantum statistical physics. Extensive interests in these systems
arise from many fields of physics, e.g., condensed matter physics
\cite{cald}, quantum optics \cite{Gar,Wil,walls}, quantum
measurement \cite{Zur,Wal}, quantum computation \cite{Unr,Lan}, and
so on. The dissipation and decoherence are generated due to the
interaction between the system and the thermal bath or reservoir.
After the enormous irrelevant degrees of freedom of the thermal bath
are integrated out from the von Nuemann equation of the density
matrix of the extended system including the environment, the master
equation for the reduced density matrix of the relevant system is
resulted in some reasonable approximations; for details see, for
example, Ref. \cite{Gar,Wil}). For a simple system such as a
two-level atom interacting with the background thermal radiation
field, as the parameters of the master equation are independent of
time, the solutions have been obtained and studied very well
\cite{Gar,Wil,walls}. However, for the non-autonomous case where the
parameters of the master equation are dependent on time, Ellinas et
al. have studied the adiabatic evolution and the corresponding Berry
phase in optical resonance \cite{ellinas}, and the problem for the
general solutions still remains open. For more complicated systems,
the solution to the master equation of the reduced density matrix
is, in general, difficult to obtain. To solve the problem, the usual
way is to convert the master equation to a set of differential
equations for some quantum statistical moments or expansion
coefficients in terms of some bases truncated at a reasonable order
\cite{Gar,walls}. In the meanwhile, some other useful approximate
methods have also been proposed, for instance, the short-time expansion \cite%
{ji}, the small lose rate expansion \cite{Yi}, the stochastic unraveling
\cite{plen}, and finally numerical calculations \cite{Wik}, etc. Some exact
methods have been also explored \cite{ven,yi2}. An elegant method was
proposed by Briegel and Englert to treat the quantum optical master
equations by using the damping bases \cite{Briegel-Englert}, but the problem
was restricted to the autonomous case where the parameters of the master
equation are independent of time.

In this paper we shall present a novel algebraic dynamic method to solve the
master equations which in many cases are found to have some dynamical
algebraic structures. The common feature of the quantum master equations is
the existence of the sandwich terms of the Liouville operators where the
reduced density matrix of the system is in between some quantum excitation
and de-excitation operators, which come from the elimination of the
environment degrees of freedom. The sandwich structure of the quantum master
equation also appears in thermal field theory where the so-called thermal
Lie algebra has been proposed to treat the problem \cite{santano, takahashi}%
, but the parameters of the master equation are still time-independent.

Our new algebraic method is just a generalization of the algebraic dynamical
method \cite{wang93} from quantum mechanical systems to quantum statistical
systems with time-dependent parameters. It is designed to treat the sandwich
problem and to explore the dynamical algebraic structures of the master
equations with time-dependent parameters built in. To this end, the right
and left representations \cite{wang89} as well as the adjoint
representations of dynamical algebras are developed and used. This new
method has been used successfully to solve the von Nuemann equation for the
quantum statistical characteristic function of the two-level Janes-Cummings
model \cite{wang01} and the master equation for the sympathetic cooling of
the Bose-Einstein condensate system in the mean field approximation \cite%
{wang02}. In this paper, we shall apply this method to solve the
master equation of the dissipative two-level atom system in the
non-autonomous case, and to study the decoherence of the multiple
two level atom system coupled to the radiation thermal bath, which
have not been studied before for the non-autonomous case.

The paper is organized as follows. In Sect. II, the model Hamiltonian of the
system is given and the master equation for the reduced density matrix of
the atom is described. In Sect. III, the dynamical u(2) algebraic structure
of the Liouville or rate operator of the master equation is found by
introducing the new composite algebras which are constructed from the right
and left representations of the relevant algebras, and the dynamical
symmetry of the system is established and the integrability of the master
equation is thus proved by using the algebraic dynamical theorem \cite%
{wang93}. Sect. IV is devoted to the analytical solutions of the
master equation for the non-autonomous case where the parameters of
the rate operator (or Hamiltonian)are dependent on time, and the
approach to the steady solution asymptotically is thus proved. In
Sect. V, the dissipation and decoherence of the multiple two-level
atom systems are investigated for the non-autonomous case.
Discussions and conclusions are given in the final section.

\section{Dissipative two-level atom system in non-autonomous case}

Consider the two-level atom interacting with a thermal quantum radiation
field. With the dipole interaction and in the rotating wave approximation,
the total system is described by the following model Hamiltonian,
\begin{equation}
\hat{H}=\frac{1}{2}\hbar \omega _{0}\sigma _{z}+\hat{H}_{bath}+\hbar (\sigma
_{+}\hat{\Lambda}+\sigma _{-}\hat{\Lambda}^{\dagger }),  \label{Hami}
\end{equation}%
where $\hat{H}_{bath}=\sum_{k}\hbar \omega _{k}b_{k}^{+}b_{k}$ describing
the background quantum radiation field, $\hat{\Lambda}=\sum_{k}g_{k}b_{k}$
is an operator used to describe the coupling between the atom and the
radiation field with coupling constants $g_{k}$. Here $\omega _{0}$ and $%
\omega _{k}$ are the transition frequency between two levels of the atom and
the mode frequencies of the radiation field, respectively. $\sigma _{+}$, $%
\sigma _{-}$, and $\sigma _{z}$ are dimensionless atomic operators obeying
the usual Pauli matrix commutation relations \cite{Dic}. Using the standard
technique from quantum optics, one obtains the master equation for the
reduced density matrix of the atom \cite{walls}
\begin{eqnarray}
\frac{\mathrm{{d}\hat{\rho}(t)}}{\mathrm{{d}t}} &=&-\frac{i}{2}\omega
_{0}[\sigma _{z},\hat{\rho}(t)]-\frac{\gamma }{2}\ (\bar{n}_{0}+1)\ (\sigma
_{+}\sigma _{-}\hat{\rho}(t)+\hat{\rho}(t)\sigma _{+}\sigma _{-}-2\sigma _{-}%
\hat{\rho}(t)\sigma _{+})  \nonumber \\
&&-\frac{\gamma }{2}\ \bar{n}_{0}\ (\sigma _{-}\sigma _{+}\hat{\rho}(t)+\hat{%
\rho}(t)\sigma _{-}\sigma _{+}-2\sigma _{+}\hat{\rho}(t)\sigma _{-})
\label{sch}
\end{eqnarray}%
where $\bar{n}_{0}$ is the mean number of photons in the environment and $%
\gamma $ denotes the damping rate. They read
\begin{eqnarray}
&&\bar{n}_{0}=[\exp (\hbar \omega _{0}/k_{B}T)-1]^{-1},  \nonumber \\
&&\gamma =2\pi \sum_{k}g_{k}^{2}\delta (\omega _{0}-\omega _{k})  \label{par}
\end{eqnarray}%
Here the term which gives rise to a small Lamb frequency shift
$\Delta \omega $ has been neglected. Equation (\ref{sch}) describes
an atom interacting with a thermal field at the temperature $T$. If
$T=0$, then $ \bar{n}_{0}=0$.

In the autonomous case, the system has been studied very well.
However, it is also interesting to control the system through
changing the temperature of the thermal bath, the atomic energy
level and the coupling constant. In this case, the parameters of the
rate operator $\gamma ,\bar{n}_{0},$ and $ \omega _{0}$ are
time-dependent, and the system becomes non-autonomous. More
basically, even if the total Hamiltonian of the composite system-the
system to be investigated plus the environment, is autonomous, the
master equation of the reduced density matrix of the investigated
system still becomes non-autonomous under the non-Markovian dynamics
\cite{anastopoulos}. Therefore, the quantum master equation of the
reduced density matrix, in general and more rigorously, should be
non-autonomous in the sense that its parameters should be
time-dependent. To our knowledge, the problem of the non-autonomous
systems has not been solved till now. Therefore, in the present
paper it is our main goal to solve the problem of the non-autonomous
case for the two-level system. In the next section we first study
the algebraic structure of the master equation(\ref{sch}) and
explore its dynamical symmetry.

\section{Algebraic structure of the master equation}

\subsection{\protect\bigskip Right and left algebras in the von Neumann space%
}

Following the idea of Ref. \cite{wang02}, the right and left algebras are
introduced. First, one notices that the density matrix $\hat{\rho}$ is a
super vector in the von Neumann space \cite{wang89},
\begin{equation}
\hat{\rho}=\sum_{s,s^{\prime }}\rho _{ss^{\prime }}|s\rangle \langle
s^{\prime }|  \label{rou}
\end{equation}%
where $|s\rangle $ denotes the Fermion state. The $\sigma _{+}$, $\sigma
_{-} $, and $\sigma _{z}$ can operate on the ket state $|s\rangle $ to the
right and on the bra state $\langle s|$ to the left, which form the right
and left representations of the usual $su(2)$ algebra as follows~\cite%
{wang89}
\begin{equation}
su(2)_{R}:\{\sigma _{z}^{r},\sigma _{+}^{r},\sigma _{-}^{r}\},\;\;\
su(2)_{L}:\{\sigma _{z}^{l},\sigma _{+}^{l},\sigma _{-}^{l}\}.  \label{alg}
\end{equation}%
They obey the commutation relations respectively as follows,
\begin{eqnarray}
&&[\sigma _{z}^{r},\sigma _{\pm }^{r}]=\pm 2\sigma _{\pm }^{r},\;\;[\sigma
_{+}^{r},\sigma _{-}^{r}]=\sigma _{z}^{r},  \nonumber \\
&&[\sigma _{z}^{l},\sigma _{\pm }^{l}]=\mp 2\sigma _{\pm }^{l},\;\;[\sigma
_{+}^{l},\sigma _{-}^{l}]=-\sigma _{z}^{l}.  \label{comm}
\end{eqnarray}%
It is evident that $su(2)_{R}$ is isomorphic to the $su(2)$, while $%
su(2)_{L} $ is anti-isomorphic to the $su(2)$. This is because that $%
su(2)_{R}$ operates, as usual, towards the right on $|s\rangle $. On the
other hand, the $su(2)_{L}$ operates towards the left on $\langle s|$. Since
$su(2)_{R}$ and $su(2)_{L}$ operate on different spaces(the ket and the bra
spaces), they commute each other, i.e.
\begin{equation}
\lbrack su(2)_{L},su(2)_{R}]=0.  \label{lr}
\end{equation}%
\ \ \ \ \ \ \

\subsection{\protect\bigskip Composite algebra and algebraic structure of
the master equation}

After having introduced the left and right algebras in Eqs. (\ref{alg}), one
can construct the composite $su(2)$ and $u(1)$ algebras as follows,
\begin{eqnarray}
su(2) &:&\{\hat{J}_{0}=\frac{\sigma _{z}^{r}+\sigma _{z}^{l}}{2},\hat{J}%
_{+}=\sigma _{+}^{r}\sigma _{-}^{l},\hat{J}_{-}=\sigma _{-}^{r}\sigma
_{+}^{l}\}.  \nonumber \\
u(1) &:&\hat{U}_{0}=\frac{\sigma _{z}^{r}-\sigma _{z}^{l}}{2}  \label{comp}
\end{eqnarray}%
According to Eqs. (\ref{comm}) it is easy to check the following commutation
relations
\begin{eqnarray}
&&[\hat{J}_{0},J_{\pm }]=\pm 2J_{\pm },\;\;[\hat{J}_{+},\hat{J}_{-}]=\hat{J}%
_{0},  \nonumber \\
&&[\hat{U}_{0},\hat{J}_{\pm }]=0,\;\;\;[U_{0},\hat{J}_{0}]=0.  \label{Re}
\end{eqnarray}%
The action of the composite $su(2)$ and $u(1)$ algebras on the bases of
von-Neumann space is
\begin{eqnarray}
\hat{J}_{0}|s\rangle \langle s^{\prime }| &=&\frac{s+s^{\prime }}{2}%
|s\rangle \langle s^{\prime }|,  \nonumber \\
\hat{J}_{+}|s\rangle \langle s^{\prime }| &=&\delta _{s+1,0}\delta
_{s^{\prime }+1,0}|s+2\rangle \langle s^{\prime }+2|,  \nonumber \\
\hat{J}_{-}|s\rangle \langle s^{\prime }| &=&\delta _{s-1,0}\delta
_{s^{\prime }-1,0}|s-2\rangle \langle s^{\prime }-2|,  \nonumber \\
\hat{U}_{0}|s\rangle \langle s^{\prime }| &=&\frac{s-s^{\prime }}{2}%
|s\rangle \langle s^{\prime }|.  \label{act}
\end{eqnarray}%
where $s$ and $s^{\prime }$ are equal to -1 or +1.

Noticing the following identities
\begin{equation}
\sigma _{+}\sigma _{-}=\frac{1+\sigma _{z}}{2},\;\;\;\ \sigma _{-}\sigma
_{+}=\frac{1-\sigma _{z}}{2},  \label{ide}
\end{equation}%
and the composite algebra Eqs. (\ref{comp}) introduced above, the master
equation(\ref{sch}) can be rewritten as
\begin{equation}
\frac{\mathrm{{d}\hat{\rho}(t)}}{\mathrm{{d}t}}=\hat{\Gamma}(t)\hat{\rho}(t),
\label{mast}
\end{equation}%
where the rate operator $\hat{\Gamma}$ reads
\begin{equation}
\hat{\Gamma}=-i\omega _{0}(t)\hat{U}_{0}+\gamma (t)\bar{n}_{0}(t)\hat{J}%
_{+}+\gamma (t)[\bar{n}_{0}(t)+1]\hat{J}_{-}-\frac{1}{2}\gamma (t)\hat{J}%
_{0}-\frac{1}{2}\gamma (t)[2\bar{n}_{0}(t)+1]  \label{rat}
\end{equation}%
The master equation (\ref{mast}) is now put into a form similar to the
time-dependent Schr\"{o}dinger equation except that the imaginary number $%
"i" $ is missing from the left hand side and the rate operator $\hat{\Gamma}$
is non-Hermitian, indicating the dissipative behavior of the system due to
the energy exchange with the thermal bath. In Eq. (\ref{mast}), the rate
operator $\hat{\Gamma}$ plays the role of the Hamiltonian and the reduced
density matrix plays the role of the wave function. Since the rate operator $%
\hat{\Gamma}$ is a linear function of the $su(2)\oplus u(1)$ generators, the
master equation (\ref{mast}) possesses the $su(2)\oplus u(1)$ dynamical
symmetry; it is integrable and can be solved analytically according to the
algebraic dynamics \cite{wang93}.

\section{Exact solution to the master equation in non-autonomous case}

\subsection{Eigensolutions of the rate operators $\hat{\Gamma}$}

In order to better understand the time-dependent solution of the master
equation, its decay behavior and its approach to the equilibrium state, we
first consider the steady eigenvalue problem of the rate operator $\hat{%
\Gamma}$ whose eigensolution itself is interesting and peculiar, and
contains the steady equilibrium state. The eigen equation reads
\begin{equation}
\hat{\Gamma}\rho (s,s^{\prime })=\beta (s,s^{\prime })\rho (s,s^{\prime }),
\label{eigen}
\end{equation}%
where $\beta (s,s^{\prime })$ is the eigenvalue of the rate operator $\hat{%
\Gamma}$ and $(s,s^{\prime })$ label the eigenstates in the von Neumann
space. This eigenvalue equation is time-independent and can be solved by
introducing the following similarity transformation
\begin{equation}
\rho (s,s^{\prime })=\hat{U}\bar{\rho}(s,s^{\prime }),  \label{simi}
\end{equation}%
where
\begin{equation}
\hat{U}=e^{\alpha _{+}\hat{J}_{+}}e^{\alpha _{-}\hat{J}_{-}},\;\;\hat{U}%
^{-1}=e^{-\alpha _{-}\hat{J}_{-}}e^{-\alpha _{+}\hat{J}_{+}}.  \label{tran}
\end{equation}%
Here $\alpha _{\pm }$ are the parameters specifying the similarity
transformation. After some calculations, one has the transformed eigenvalue
equation as follows,
\begin{eqnarray}
&&\bar{\Gamma}\bar{\rho}(s,s^{\prime })=\beta (s,s^{\prime })\bar{\rho}%
(s,s^{\prime }),  \nonumber \\
&&\bar{\Gamma}=\hat{U}^{-1}\hat{\Gamma}\hat{U},  \label{dia}
\end{eqnarray}%
where the transformed rate operator $\bar{\Gamma}$ is diagonalized and
becomes a linear combination of the commuting invariant operators $\hat{J}%
_{0}$ and $\hat{U}_{0}$ which dictate the dynamical symmetry of the system,
\begin{equation}
\bar{\Gamma}=-i\omega _{0}\hat{U}_{0}-\frac{1}{2}\gamma \lbrack 2(\bar{n}%
_{0}+1)\alpha _{+}+1]\hat{J}_{0}-\frac{1}{2}\gamma (2\bar{n}_{0}+1)
\label{bar}
\end{equation}%
if the following diagonalization conditions are fulfilled
\begin{eqnarray}
&&-(\bar{n}_{0}+1)\alpha _{+}^{2}-\alpha _{+}+\bar{n}_{0}=0,  \nonumber \\
&&(\bar{n}_{0}+1)(1+2\alpha _{+}\alpha _{-})+\alpha _{-}=0.  \label{cond}
\end{eqnarray}%
The eigenvectors of $\bar{\Gamma}$ are the common solutions of $\hat{J}_{0}$
and $\hat{U}_{0}$, just the form of $|s\rangle \langle s^{\prime }|$ with $s$
and $s^{\prime }=\pm 1.$ Eqs. (\ref{cond}) have two sets of solutions, which
yield two sets of eigenvalues $\beta (s,s^{\prime })$ for the rate operator
\begin{eqnarray}
(a) &&\;\;\alpha _{+}=-1,\;\;\alpha _{-}=\frac{\bar{n}_{0}+1}{2\bar{n}_{0}+1}%
,\;\;\beta (s,s^{\prime })=-i\omega _{0}\frac{s-s^{\prime }}{2}+\frac{\gamma
}{2}(2\bar{n}_{0}+1)(\frac{s+s^{\prime }}{2}-1),  \nonumber \\
(b) &&\;\;\alpha _{+}=\frac{\bar{n}_{0}}{\bar{n}_{0}+1},\;\;\alpha _{-}=-%
\frac{\bar{n}_{0}+1}{2\bar{n}_{0}+1},\;\;\beta (s,s^{\prime })=-i\omega _{0}%
\frac{s-s^{\prime }}{2}+\frac{\gamma }{2}(2\bar{n}_{0}+1)(-\frac{s+s^{\prime
}}{2}-1).  \label{sol}
\end{eqnarray}%
At first glance, it is surprising that two similarity transformations exist
to diagonalize the same rate operator and to yield two sets of eigenvalues.
This is in contrast to the diagonalization of a Hamiltonian where the
unitary transformation to diagonalize the Hamiltonian is usually unique and
the set of the eigensolutions is also unique. The peculiar results stem from
the special structure of the rate operator (\ref{rat}): (1) it contains a
part(from the second to the forth terms) which is a vector in the linear
space spanned by the $su(2)$ generators($\hat{J}_{+},\hat{J}_{-},\hat{J}_{0}$%
), and allows two transformations to rotate this part of vector along the $%
\hat{J}_{0}$ and $-\hat{J}_{0}$ directions; (2) the last term of the rate
operator is a constant term which is a scalar in the $su(2)$ space and make
above two diagonalizing transformation asymmetric. The above two features
result in two similarity transformations \cite{wang02}. However, as will be
seen soon, after returning to the physical frame by the inverse
transformations, the two sets of eigensolutions coincide. This means that
the physical results are objective, independent of the similarity
transformations used.

It is interesting to note that both solution $(a)$ and $(b)$ contain the
zero-mode steady solution and the nonzero- mode decaying solutions( with
negative eigenvalues ), which guarantee that any time-dependent solution of
the reduced density matrix asymptotically approaches the steady solution as
shown below.

Performing an inverse transformation, the eigensolutions of the rate
operator is obtained readily
\begin{equation}
\rho (s,s^{\prime })=\hat{U}\bar{\rho}(s,s^{\prime })=(1+\alpha
_{+}J_{+})(1+\alpha _{-}J_{-})|s\rangle \langle s^{\prime }|.  \label{inver}
\end{equation}%
Explicitly, both (a) and (b) solutions lead to the same physical
eigensolutions.
\begin{eqnarray}
\beta _{1} &=&\beta (-1,-1)=0,\text{ \ \ \ \ \ \ \ \ \ \ \ \ \ \ \ \ \ \ \ \
\ \ \ }\rho _{1}=\rho (-1,-1)=\frac{\bar{n}_{0}+1}{2\bar{n}_{0}+1}|-1\rangle
\langle -1|+\frac{\bar{n}_{0}}{2\bar{n}_{0}+1}|+1\rangle \langle +1|,
\nonumber \\
\beta _{2} &=&\beta (+1,+1)=-\gamma (2\bar{n}_{0}+1),\text{ \ \ \ \ \ \ \ \
\ \ \ \ \ }\rho _{2}=\rho (+1,+1)=|-1\rangle \langle -1|-|+1\rangle \langle
+1|,  \nonumber \\
\beta _{3} &=&\beta (+1,-1)=-\frac{\gamma }{2}(2\bar{n}_{0}+1)-i\omega _{0},%
\text{ \ \ \ \ \ }\rho _{3}=\rho (+1,-1)=|+1\rangle \langle -1|,  \nonumber
\\
\beta _{4} &=&\beta (-1,+1)=-\frac{\gamma }{2}(2\bar{n}_{0}+1)+i\omega _{0},%
\text{ \ \ \ \ \ }\rho _{4}=\rho (-1,+1)=|-1\rangle \langle +1|.  \label{fou}
\end{eqnarray}%
where the first line of Eqs. (\ref{fou}) is the zero-mode solution
corresponding to the steady state.

Another feature of the rate operator is its non-Hermiticity, i.e., $\hat{%
\Gamma}^{\dagger }\neq \hat{\Gamma}$, which is evident from $\hat{J}%
_{+}^{\dagger }=\hat{J}_{-}$, $\hat{J}_{-}^{\dagger }=\hat{J}_{+}$, $\hat{J}%
_{0}^{\dagger }=\hat{J}_{0},$ and $\hat{U}_{0}^{\dagger }=\hat{U}_{0}$.
Since $\hat{\Gamma}$ is non-Hermitian, the eigenvectors of $\hat{\Gamma}$
and $\hat{\Gamma}^{\dagger }$ constitute a bi-orthogonal basis \cite{Mor}.

Using a similarity transformation $\hat{U}^{^{\prime }}=e^{-\alpha _{+}\hat{J%
}_{-}}e^{-\alpha _{-}\hat{J}_{+}}$ and under the conditions of Eq. (\ref%
{cond}), diagonalization of the operator $\hat{\Gamma}^{\dagger }$ can be
obtained as follows
\begin{equation}
\bar{\Gamma}^{\dagger }=i\omega _{0}\hat{U}_{0}-\frac{1}{2}\gamma \lbrack 2(%
\bar{n}_{0}+1)\alpha _{+}+1]\hat{J}_{0}-\frac{1}{2}\gamma (2\bar{n}_{0}+1)
\label{dg}
\end{equation}%
In a similar way, we get the eigensolutions of $\hat{\Gamma}^{\dagger }$.
\begin{eqnarray}
\beta _{1}^{\ast } &=&\beta ^{\ast }(-1,-1)=0,\text{ \ \ \ \ \ \ \ \ \ \ \ \
\ \ \ \ \ \ }\tilde{\rho}_{1}=\tilde{\rho}(-1,-1)=|-1\rangle \langle
-1|+|+1\rangle \langle +1|,  \nonumber \\
\beta _{2}^{\ast } &=&\beta^{\ast }(+1,+1)=-\gamma (2\bar{n}_{0}+1),\text{ \
\ \ \ \ \ \ \ }\tilde{\rho}_{2}=\tilde{\rho}(+1,+1)=\frac{\bar{n}_{0}}{2\bar{%
n}_{0}+1}|-1\rangle \langle -1|-\frac{\bar{n}_{0}+1}{2\bar{n}_{0}+1}%
|+1\rangle \langle +1|,  \nonumber \\
\beta _{3}^{\ast } &=&\beta^{\ast }(+1,-1)=-\frac{\gamma }{2}(2\bar{n}%
_{0}+1)+i\omega _{0},\text{\ }\tilde{\rho}_{3}=\tilde{\rho}%
(+1,-1)=|+1\rangle \langle -1|,  \nonumber \\
\beta _{4}^{\ast } &=&\beta^{\ast }(-1,+1)=-\frac{\gamma }{2}(2\bar{n}%
_{0}+1)-i\omega _{0},\text{\ }\tilde{\rho}_{4}=\tilde{\rho}%
(-1,+1)=|-1\rangle \langle +1|.  \label{foudg}
\end{eqnarray}%
where $\beta _{j}^{\ast }$ and $\tilde{\rho}_{j}$ are the eigenvalues and
eigenvectors of the operator $\hat{\Gamma}^{\dagger }$. It is can be checked
that Eqs. (\ref{fou}) and Eqs. (\ref{foudg}) are bi-orthogonal.

\subsection{Time-dependent solutions of the master equation in Non-autonomous
case}

Since Eq. (\ref{rat}) is a linear combination of the generators of the
composite algebra, the master equation Eq. (\ref{mast}) possesses the $%
su(2)\oplus u(1)$ dynamical symmetry and is thus integrable,
solvable analytically even in the non-autonomous case \cite{wang93}.

The master equation in the non-autonomous case can be solved by
algebraic dynamical method via the following time-dependent gauge
transformation which \ is a generalization of time-independent
similarity transformations in the autonomous case to the
non-autonomous case ( the terminology of gauge transformation is due
to the fact that it induces a gauge term in the rate operator
similar to the gauge field theory),
\begin{equation}
\hat{U}_{g}=e^{\alpha _{+}(t)\hat{J}_{+}}e^{\alpha _{-}(t)\hat{J}_{-}}.
\label{gauge}
\end{equation}%
After the gauge transformation and under the following gauge conditions
\begin{eqnarray}
&&\frac{d\alpha _{+}(t)}{dt}=-\gamma (t)[\bar{n}_{0}(t)+1]\alpha
_{+}^{2}(t)-\gamma (t)\alpha _{+}(t)+\gamma (t)\bar{n}_{0}(t)=-\gamma (t)[%
\bar{n}_{0}(t)+1][\alpha _{+}(t)+1][\alpha _{+}(t)-\frac{\bar{n}_{0}(t)}{%
\bar{n}_{0}(t)+1}]  \nonumber \\
&&\frac{d\alpha _{-}(t)}{dt}=\gamma (t)[\bar{n}_{0}(t)+1][1+2\alpha
_{+}(t)\alpha _{-}(t)]+\gamma (t)\alpha _{-}(t),  \label{diff}
\end{eqnarray}%
\ the gauged rate operator is diagonalized and the gauged master equation
becomes simple and integrable,
\begin{eqnarray}
\frac{d\bar{\rho}(t)}{dt} &=&\bar{\Gamma}(t)\bar{\rho}(t),  \nonumber \\
\bar{\Gamma}(t) &=&\hat{U}_{g}^{-1}\Gamma (t)\hat{U}_{g}-\hat{U}_{g}^{-1}%
\frac{d\hat{U}_{g}}{dt}=-i\omega _{0}(t)\hat{U}_{0}-\frac{1}{2}\gamma (t)\{2[%
\bar{n}_{0}(t)+1]\alpha _{+}(t)+1]\hat{J}_{0}-\frac{1}{2}\gamma (t)[2\bar{n}%
_{0}(t)+1]\}.  \label{j}
\end{eqnarray}%
which is a linear function of the complete set of the commuting operators
(invariant operators)$\hat{U}_{0}$ and $\hat{J}_{0}$, and clearly shows a $%
u(2)$ dynamical symmetry. The solution of Eqs. (\ref{j}) reads
\begin{equation}
\bar{\rho}(t)=e^{\int_{0}^{t}\bar{\Gamma}(\tau )d\tau }\bar{\rho}(0),
\label{rob}
\end{equation}

For the initial conditions
\begin{eqnarray}
\alpha _{+}(0) &=&\alpha _{-}(0)=0,  \nonumber \\
or\text{ }\rho (0) &=&\bar{\rho}(0)=\sum_{ss^{\prime }}p_{ss^{\prime
}}|s\rangle \langle s^{\prime }|
\end{eqnarray}%
or , we finally obtain the solution,
\begin{equation}
\rho (t)=e^{\alpha _{+}(t)\hat{J}_{+}}e^{\alpha _{-}(t)\hat{J}_{-}}e^{-i\hat{%
U}_{0}\int_{0}^{t}\omega _{0}(\tau )d\tau -\hat{J}_{0}\int_{0}^{t}\gamma
(\tau )((\bar{n}_{0}(\tau )+1)\alpha _{+}(\tau )+\frac{1}{2})d\tau
-\int_{0}^{t}\frac{\gamma (\tau )}{2}(2\bar{n}_{0}(\tau )+1)d\tau }\rho (0).
\label{rout2}
\end{equation}

Once the reduced density matrix of the non-autonomous system is
obtained, the
averages of the physical observables $\sigma _{z}$, $\sigma _{+}$, and $%
\sigma _{+}$ can be calculated. For the system initially in a pure state, we
have $\rho (0)=|\mu |^{2}|1\rangle \langle 1|+|\nu |^{2}|-1\rangle \langle
-1|+\mu \nu ^{\ast }|1\rangle \langle -1|+\mu ^{\ast }\nu |-1\rangle \langle
1|$, where $|\mu |^{2}+|\nu |^{2}=1$. Then one obtains
\begin{eqnarray}
\rho (t) &=&[f_{1,1}(t)|\mu |^{2}(1+\alpha _{+}(t)\alpha
_{-}(t))+f_{-1,-1}(t)|\nu |^{2}\alpha _{+}(t)]|1\rangle \langle
1|+[f_{1,1}(t)|\mu |^{2}\alpha _{-}(t)+f_{-1,-1}(t)|\nu |^{2}]|-1\rangle
\langle -1|  \nonumber \\
&&+f_{1,-1}(t)\mu \nu ^{\ast }|1\rangle \langle -1|+f_{-1,1}(t)\mu ^{\ast
}\nu |-1\rangle \langle 1|,  \label{rout3}
\end{eqnarray}%
where
\begin{equation}
f_{s,s^{\prime }}(t)=e^{-i\frac{s-s^{\prime }}{2}\int_{0}^{t}\omega
_{0}(\tau )d\tau -\frac{s+s^{\prime }}{2}\int_{0}^{t}\gamma (\tau )\{[\bar{n}%
_{0}(\tau )+1]\alpha _{+}(\tau )+\frac{1}{2}\}d\tau -\frac{1}{2}%
\int_{0}^{t}\gamma (\tau )[2\bar{n}_{0}(\tau )+1]d\tau },  \label{f}
\end{equation}%
From Eq. (\ref{rout3}), we get
\begin{eqnarray}
\langle \sigma _{z}\rangle &=&f_{1,1}(t)|\mu |^{2}[1+\alpha _{+}(t)\alpha
_{-}(t)-\alpha _{-}(t)]+f_{-1,-1}(t)|\nu |^{2}[\alpha _{+}(t)-1],  \nonumber
\\
\langle \sigma _{+}\rangle &=&f_{-1,1}(t)\mu ^{\ast }\nu  \nonumber \\
\langle \sigma _{-}\rangle &=&f_{1,-1}(t)\mu \nu ^{\ast }.  \label{mea3}
\end{eqnarray}

For the autonomous case, $\bar{n}_{0}$ and $\gamma $ are independent of
time, $\alpha _{+}$ and $\alpha _{-}$ have the following analytical
solutions,
\begin{eqnarray}
\alpha _{+}(t) &=&\frac{1-e^{-\gamma (2\bar{n}_{0}+1)t}}{\frac{\bar{n}_{0}+1%
}{\bar{n}_{0}}+e^{-\gamma (2\bar{n}_{0}+1)t}},  \nonumber \\
\alpha _{-}(t) &=&\frac{(\bar{n}_{0}+1)\bar{n}_{0}[\frac{\bar{n}_{0}+1}{\bar{%
n}_{0}}+e^{-\gamma (2\bar{n}_{0}+1)t}][1-e^{-\gamma (2\bar{n}_{0}+1)t}]}{(2%
\bar{n}_{0}+1)^{2}e^{-\gamma (2\bar{n}_{0}+1)t}}.  \label{af}
\end{eqnarray}%
Then
\begin{eqnarray}
f_{1,1}(t) &=&\frac{(2\bar{n}_{0}+1)e^{-\gamma (2\bar{n}_{0}+1)t}}{(\bar{n}%
_{0}+1)+\bar{n}_{0}e^{-\gamma (2\bar{n}_{0}+1)t}},  \nonumber \\
f_{-1,-1}(t) &=&\frac{(\bar{n}_{0}+1)+\bar{n}_{0}e^{-\gamma (2\bar{n}%
_{0}+1)t}}{2\bar{n}_{0}+1},  \nonumber \\
f_{1,-1}(t) &=&e^{-i\omega _{0}t-\frac{\gamma }{2}(2\bar{n}_{0}+1)t},
\nonumber \\
f_{-1,1}(t) &=&e^{i\omega _{0}t-\frac{\gamma }{2}(2\bar{n}_{0}+1)t}.
\label{f2}
\end{eqnarray}%
Inserting Eqs. (\ref{af}) and Eqs. (\ref{f2}) into Eqs. (\ref{mea3}), we
recover the well-known results for the autonomous system.

\subsection{Approaching the steady solution}

Like the autonomous case, the time-dependent solutions of the master
equation in the non-autonomous case also asymptotically approaches
the steady solution satisfying
\begin{equation}
\frac{\mathrm{{d}\hat{\rho}(t)}}{\mathrm{{d}t}}=\hat{\Gamma}\hat{\rho}(t)=0
\label{zero2}
\end{equation}%
which has the same solution as the zero-mode eigensolutions of $\hat{\Gamma}$
, namely the steady solutions with the parameters $\alpha _{+}$ and $\alpha
_{-}$ obeying
\begin{eqnarray}
&&\frac{d\alpha _{+}(t)}{dt}=0,  \nonumber \\
&&\frac{d\alpha _{-}(t)}{dt}=0.  \label{diff2}
\end{eqnarray}%
whose solutions are obviously the same as Eqs. (\ref{cond}) and
should have the two same sets of solutions $(a)$ and $(b)$ as given
in Eqs. (\ref{sol}). For the autonomous case, the two similarity
transformations as tools to diagonalize the rate operator are on the
equal footing and generate the same physical solution as proved
above; while for the non-autonomous case, the properties of the two
sets of solutions of $\alpha _{+}$ and $\alpha _{-}$ should be
examined on the background of their time evolution. It is found from
the Eq. (26) that if $\alpha _{+}=-1-\epsilon ,$then $\frac{d\alpha
_{+}(t)}{dt}<0$ and $\alpha _{+}$ will go away further from $-1$
towards negative direction; while as $\alpha _{+}=-1+\epsilon $,
then $\frac{d\alpha _{+}(t)}{dt}>0$ and $\alpha _{+}$ will go away
further from $-1$ towards positive direction. Thus the steady
solution $\alpha _{+}=-1$ is unstable. Since the solution of $\alpha
_{-}(t)$ depends on $\alpha _{+}(t),$ it is also unstable. Therefore
the solution (a) is unstable and can not be reached from the initial
condition $\alpha _{+}=\alpha _{-}=0$. Instead, the solution (b) is
partly stable in the following sense. Because the first equation of
Eq. (26) reads
\begin{equation}
\frac{d\alpha _{+}(t)}{dt}=-\gamma (t)[\bar{n}_{0}(t)+1][\alpha
_{+}(t)+1][\alpha _{+}(t)-\frac{\bar{n}_{0}(t)}{\bar{n}_{0}(t)+1}].
\end{equation}
It's easy to see that
\[
\left\{
\begin{array}{c}
d\alpha _{+}(t)/dt>0,\text{ if }0<d\alpha _{+}(t)<\frac{\bar{n}_{0}(t)}{\bar{%
n}_{0}(t)+1} \\
d\alpha _{+}(t)/dt<0,\text{\ if\ }\alpha _{+}(t)>\frac{\bar{n}_{0}(t)}{\bar{n%
}_{0}(t)+1}\text{ and }\alpha _{+}(t)<-1%
\end{array}%
\right.
\]%
$d\alpha _{+}(t)/dt>0(<0)$ if $d\alpha _{+}(t)/dt>0$ (if $\alpha _{+}(t)>%
\frac{\bar{n}_{0}(t)}{\bar{n}_{0}(t)+1}$ and $\alpha _{+}(t)<-1$). With the
initial condition $\alpha _{+}=0$, we see that $\alpha _{+}$ approaches the
value $\frac{\bar{n}_{0}(\infty )}{\bar{n}_{0}(\infty )+1}=\frac{\bar{n}_{0}%
}{\bar{n}_{0}+1}$ asymptotically from zero. However, $\alpha _{-}(t)$ can
not reach its steady value $-\frac{\bar{n}_{0}+1}{2\bar{n}_{0}+1}$. To study
the asymptotic behavior of $\alpha _{-}$, we define $y(t)=\alpha
_{-}(t)\times \exp \{-\int_{0}^{t}\gamma (\tau )(\bar{n}_{0}(\tau
)+1)(\alpha _{+}(\tau )+1)d\tau \}=\alpha _{-}(t)\exp \int_{0}^{t}p(\tau
)d\tau $. The time differential of $y\left( t\right) $ is given by $b\exp
\int_{0}^{t}p(\tau )d\tau $ where $b=\dot{\alpha}_{-}(t)+\alpha _{-}(t)p(t)$%
. Since $b\longrightarrow \gamma (\bar{n}_{0}+1)$ which is bounded and $p(t)$
is negative for large $t$, the differential tends to zero and, hence, $%
y\left( t\right) $ is towards a constant. This implies that $\alpha _{-}(t)$
diverges asymptotically. So the unique solution of Eq. (26) has the
following asymptotic properties:
\begin{eqnarray}
&&\alpha _{+}(\infty )=\frac{\bar{n}_{0}}{\bar{n}_{0}+1},  \nonumber \\
&&\alpha _{-}(t)\times e^{-\int_{0}^{t}\gamma (\tau )(\bar{n}_{0}(\tau
)+1)(\alpha _{+}(\tau )+1)d\tau }\mid _{t\rightarrow \infty }=const.
\label{asy}
\end{eqnarray}%
Using the above asymptotic relations, one obtains the asymptotic results of
the time-dependent solutions as follows
\begin{eqnarray}
\rho _{+-}(t)\mid _{t\rightarrow \infty } &=&e^{-i\int_{0}^{t}\omega
_{0}(\tau )d\tau -\int_{0}^{t}(\gamma (\tau )(\bar{n}_{0}(\tau )+\frac{1}{2}%
)d\tau }e^{\alpha _{+}(t)\hat{J}_{+}}e^{\alpha _{-}(t)\hat{J}_{-}}|+1\rangle
\langle -1|  \nonumber \\
\text{ } &=&e^{-i\int_{0}^{t}\omega _{0}(\tau )d\tau -\int_{0}^{t}\gamma
(\tau )(\bar{n}_{0}(\tau )+\frac{1}{2})d\tau }|+1\rangle \langle
-1|\longrightarrow 0,  \nonumber \\
\rho _{-+}(t)\mid _{t\rightarrow \infty } &=&e^{+i\int_{0}^{t}\omega
_{0}(\tau )d\tau -\int_{0}^{t}\gamma (\tau )(\bar{n}_{0}(\tau )+\frac{1}{2}%
)d\tau }e^{\alpha _{+}(t)\hat{J}_{+}}e^{\alpha _{-}(t)\hat{J}_{-}}|-1\rangle
\langle +1|  \nonumber \\
&=&e^{+i\int_{0}^{t}\omega _{0}(\tau )d\tau -\int_{0}^{t}\gamma (\tau )(\bar{%
n}_{0}(\tau )+\frac{1}{2})d\tau }|+1\rangle \langle -1|\longrightarrow 0,
\nonumber \\
\rho _{++}(t)\mid _{t\rightarrow \infty } &=&e^{-\int_{0}^{t}\gamma (\tau )(%
\bar{n}_{0}(\tau )+1)(\alpha _{+}(\tau )+1)d\tau }e^{\alpha _{+}(t)\hat{J}%
_{+}}e^{\alpha _{-}(t)\hat{J}_{-}}|+1\rangle \langle +1|  \nonumber \\
&=&e^{-\int_{0}^{t}\gamma (\tau )(\bar{n}_{0}(\tau )+1)(\alpha _{+}(\tau
)+1)d\tau }[|+1\rangle \langle +1|+\alpha _{-}(t)(|-1\rangle \langle
-1|+\alpha _{+}(t)|+1\rangle \langle +1|)]  \nonumber \\
&\longrightarrow &const.\times (|-1\rangle \langle -1|+\frac{\bar{n}_{0}}{%
\bar{n}_{0}+1}|+1\rangle \langle +1|)=const.\times \rho _{steady},  \nonumber
\\
\rho _{--}(t)\mid _{t\rightarrow \infty } &=&e^{-\int_{0}^{t}\gamma (\tau
)\{-(\bar{n}_{0}(\tau )+1))\alpha _{+}(\tau )+\bar{n}_{0}(\tau )\}d\tau
}e^{\alpha _{+}(t)\hat{J}_{+}}e^{\alpha _{-}(t)\hat{J}_{-}}|-1\rangle
\langle -1|  \nonumber \\
&=&e^{-\int_{0}^{t}\gamma (\tau )[-(\bar{n}_{0}(\tau )+1)\alpha _{+}(\tau )+%
\bar{n}_{0}(\tau )]d\tau }[|-1\rangle \langle -1|+\alpha _{+}(t)|+1\rangle
\langle +1|]\longrightarrow c\times \rho _{steady}.  \label{tidep}
\end{eqnarray}

In the above derivation, we have used the following asymptotic relations:
\begin{equation}
e^{-\int_{0}^{t}\gamma (\tau )[-(\bar{n}_{0}(\tau )+1)(\alpha _{+}(\tau )+%
\bar{n}_{0}(\tau )]d\tau }\mid _{t\rightarrow \infty }\longrightarrow 1.
\label{p}
\end{equation}%
which can be proved by Eq. (26).

The above results indicate that the time-dependent solutions of the
master equation in the non-autonomous case also asymptotically
approach the steady solution irrespective of their initial
conditions, and that the divergent behavior of $\alpha _{-}(t)$ and
eq.(37) play an important role in the process of approaching
equilibrium state.

\section{Decoherence of multiple atom systems}

It is straightforward to generalize the above model to $N$ two-level
atoms.\ The problem of the $N$ two-level atoms system is related to
the quantum register and the entanglement state in quantum
computation \cite{zan}. \ Because of the inevitable coupling of the
atoms to the external environment, the entanglement state will lose
the coherence among different atomic states and some information
carried by the multiple atoms will be lost. Palma et al. has studied
the impact of decoherence on the efficiency of the Shor quantum
algorithm and the decoherence of quantum register at the two qubit
level \cite{palma}. However, the solution to the problem in the
non-autonomous case is still lacking. \ The $N$ two-level atoms
coupled to the quantum radiation environment can be described by the
following model Hamiltonian,
\begin{eqnarray}
\hat{H} &=&\hat{H}_{s}+\hat{H}_{env}+\hat{H}_{I},  \nonumber \\
\hat{H}_{s} &=&\frac{1}{2}\sum_{k=1}^{N}\hbar \Omega _{k}\sigma _{k}^{z},
\nonumber \\
\hat{H}_{I} &=&\sum_{k=1}^{N}\sum_{j=1}^{\infty }(g_{kj}b_{j}^{\dagger
}\sigma _{k}^{-}+h.c.).  \label{mul}
\end{eqnarray}%
Taking the same procedure and the same approximations as those for the one
atom case , the master equation for the $N$ two-level atoms can be obtained
\begin{eqnarray}
\dot{\rho}_{N} &=&\sum_{k=1}^{N}\Gamma ^{k}\rho _{N}=\Gamma \rho _{N}
\nonumber \\
\Gamma  &=&\sum_{k=1}^{N}\Gamma ^{k},  \nonumber \\
\Gamma ^{k} &=&-i\omega _{0}\hat{U}_{0}^{k}+\gamma \bar{n}_{0}\hat{J}%
_{+}^{k}+\gamma (\bar{n}_{0}+1)\hat{J}_{-}^{k}-\frac{\gamma }{2}\hat{J}%
_{0}^{k}-\frac{\gamma }{2}(2\bar{n}_{0}+1).  \label{mrou}
\end{eqnarray}%
Here we have assumed that the $N$ two level atoms are identical and coupled
to the same environment so that the decay rate $\gamma $ for different atoms
and the mean number of environment photons $\bar{n}_{0}$ are the same (if
they are different, a superscript \textquotedblright k\textquotedblright\
should be put on each pair of parameters, namely $\gamma ^{k}$ and $\bar{n}%
_{0}^{k}$ ). We have also set $\Omega _{1}=\Omega _{2}=\cdots =\Omega
_{N}=\omega _{0}$. The solution to Eqs. (\ref{mrou}) reads
\begin{equation}
\rho _{N}(t)=\prod_{k=1}^{N}\rho _{k}(t).  \label{mrouf}
\end{equation}%
According to Eq. (\ref{rout2}), the density matrix $\rho _{k}(t)$ of the
k-th qubit is
\begin{eqnarray}
\rho _{k}(t) &=&e^{\alpha _{+}^{k}(t)\hat{J}_{+}^{k}}e^{\alpha _{-}^{k}(t)%
\hat{J}_{-}^{k}}e^{\int_{0}^{t}\bar{\Gamma}^{k}(\tau )d\tau }\rho _{k}(0),
\nonumber \\
\bar{\Gamma}^{k}(t) &=&-i\omega _{0}(t)\hat{U}_{0}^{k}-\frac{1}{2}\gamma
(t)\{2[\bar{n}_{0}(t)+1]\alpha _{+}^{k}(t)+1\}\hat{J}_{0}^{k}  \nonumber \\
&&-\frac{1}{2}\gamma (t)[2\bar{n}_{0}(t)+1].  \label{mini}
\end{eqnarray}%
where $\bar{\Gamma}^{k}(t)=-i\omega _{0}(t)\hat{U}_{0}^{k}-\frac{1}{2}\gamma
(t)\{2[\bar{n}_{0}(t)+1]\alpha _{+}^{k}(t)+1\}\hat{J}_{0}^{k}-\frac{1}{2}%
\gamma (t)[2\bar{n}_{0}(t)+1].$ \ As in the last section, the initial
density matrix $\rho _{k}(0)$ of each qubit can be expanded in terms of the
superbases, $\rho _{k}(0)=\sum_{(s,s^{\prime })}c_{s,s^{\prime
}}^{k}|s\rangle ^{kk}\langle s^{\prime }|$. \ Here $\alpha _{+}^{k}(t)$ and $%
\alpha _{-}^{k}(t)$ obey the Eqs. (\ref{diff}), because we have assumed that
the parameters of the rate operators for different atoms are the same
functions of time. If $\gamma ^{k}$ and $\bar{n}_{0}^{k}$ are different for
different atoms, $\alpha _{+}^{k}(t)$ and $\alpha _{-}^{k}(t)$ are also
different for different atoms. However they obey the same form of Eqs. (\ref%
{diff}) but with different parameters $\gamma ^{k}$ and $\bar{n}_{0}^{k}$. \
To illuminate the model concretely, we also consider the two qubit system
with the initial state in the pure state, namely $|\psi (0)\rangle =\alpha
|+-\rangle +\beta |-+\rangle ,$ and%
\begin{eqnarray}
\rho _{2}(0) &=&|\alpha |^{2}|+1\rangle ^{11}\langle +1|\otimes |-1\rangle
^{22}\langle -1|+|\beta |^{2}|-1\rangle ^{11}\langle -1|\otimes |+1\rangle
^{22}\langle +1|+\alpha \beta ^{\ast }|+1\rangle ^{11}\langle -1|\otimes
|-1\rangle ^{22}\langle +1|  \nonumber \\
&&+\alpha ^{\ast }\beta |-1\rangle ^{11}\langle +1|\otimes |+1\rangle
^{22}\langle -1|  \label{xc}
\end{eqnarray}%
The time-dependent solution of the density matrix is now
\begin{eqnarray}
\rho _{2}(t) &=&|\alpha |^{2}f_{1,1}(t)[(1+\alpha _{+}(t)\alpha
_{-}(t))|+1\rangle ^{11}\langle +1|+\alpha _{-}(t)|-1\rangle ^{11}\langle
-1|]\otimes f_{-1,-1}\left( t\right) [|-1\rangle ^{22}\langle -1|+\alpha
_{+}(t)|+1\rangle ^{22}\langle +1|]  \nonumber \\
&&+|\beta |^{2}f_{-1,-1}\left( t\right) [|-1\rangle ^{11}\langle -1|+\alpha
_{+}(t)|+1\rangle ^{11}\langle +1|]\otimes f_{1,1}(t)[(1+\alpha
_{+}(t)\alpha _{-}(t))|+1\rangle ^{22}\langle +1|+\alpha _{-}(t)|-1\rangle
^{22}\langle -1|]  \nonumber \\
&&+\alpha \beta ^{\ast }f_{1,-1}|+1\rangle ^{11}\langle -1|\otimes
f_{-1,1}|-1\rangle ^{22}\langle +1|+\alpha ^{\ast }\beta f_{-1,1}|-1\rangle
^{11}\langle +1|\otimes f_{1,-1}|+1\rangle ^{22}\langle -1|  \label{rot3}
\end{eqnarray}%
where $f_{s,s^{\prime }}\left( t\right) $ are the same as given in Eqs. (\ref%
{f}). For the autonomous case, $\gamma \left( t\right) $, $\bar{n}_{0}\left(
t\right) $ and $\omega _{0}(t)$ are independent of time, and the solutions
of $\alpha _{+}(t)$ and $\alpha _{-}(t)$ are determined from Eqs. (\ref{diff}%
), which are the same as Eqs. (\ref{af}). Eq. (\ref{rot3}) is now reduced to
the result for the autonomous case, it reads

\begin{eqnarray}
\rho _{2}\left( t\right) &=&e^{-\gamma (2\bar{n}_{0}+1)t}\{2\frac{\bar{n}%
_{0}|\alpha |^{2}-(\bar{n}_{0}+1)|\beta |^{2}}{2\bar{n}_{0}+1}\rho
_{1}^{1}\otimes \rho _{2}^{2}+2\frac{\bar{n}_{0}|\beta |^{2}-(\bar{n}%
_{0}+1)|\alpha |^{2}}{2\bar{n}_{0}+1}\rho _{2}^{1}\otimes \rho
_{1}^{2}+\alpha \beta ^{\ast }\rho _{3}^{1}\otimes \rho _{4}^{2}+\alpha
^{\ast }\beta \rho _{4}^{1}\otimes \rho _{3}^{2}\}  \nonumber \\
&&-e^{-2\gamma (2\bar{n}_{0}+1)t}4\frac{\bar{n}_{0}(\bar{n}_{0}+1)}{(2\bar{n}%
_{0}+1)^{2}}\rho _{2}^{1}\otimes \rho _{2}^{2}+\rho _{1}^{1}\otimes \rho
_{1}^{2}.  \label{mrout}
\end{eqnarray}%
The above solution indicates that during the time evolution, the density
matrix of the two qubit entanglement state will approach the steady density
matrix( the last term in Eq. (\ref{mrout}) )and lose its coherence. The
characteristic time of the decoherence is $\tau _{decoh}=\frac{1}{\gamma (2%
\bar{n}_{0}+1)}$.

\section{Summary}

In this paper, Based on the quantum master equation, we have
investigated the dissipative and decoherence behaviors of the
two-level atom system coupled to the environment of thermal quantum
radiation fields. The dynamical u(2) algebraic structure of the
quantum master equation of the two-level dissipative system in the
non-autonomous case is found by virtue of left and right algebras.
By the algebraic dynamical method and proper gauge transformations,
the analytical solutions to the non-autonomous master equation are
obtained and the long time behavior of the system has been examined.
Finally we extended the model to the multiple two-level dissipative
atom system and its decoherence is studied in terms of the density
matrices for the non-autonomous case, which are given analytically
and related to quantum register and quantum computation. Since the
master equations of a wide class of dissipative quantum systems
possess some dynamical algebraic structures, the present method used
by us may serve as a useful tool in quantum statistical physics to
treat the dissipative and decoherence problems. In addition, the
results obtained in this paper may be practically useful for the
analysis of the decoherence of the multiple two-level atom systems
and quantum register.

\section{Acknowledgment}

This work was supported in part by the National Natural Science Foundation
under grants No.10175029 and 10004012, the Doctoral Education Fund of the
Education Ministry and Post-doctoral Science Foundation, and the Nuclear
Theory Research Fund of HIRFL of China.

\end{document}